# Infra-Red, *In-Situ* (IRIS) Inspection of Silicon

## Andrew 'bunnie' Huang
Sutajio Ko-Usagi Pte Ltd, Singapore
b@bunnie.org

**Abstract**

This paper introduces the Infra-Red, *In Situ* (IRIS) inspection method, which uses short-wave IR (SWIR) light to non-destructively "see through" the backside of chips and image them with lightly modified conventional digital CMOS cameras. With a ~1050 nm light source, IRIS is capable of constraining macro- and meso-scale features of a chip. This hardens existing micro-scale self-test verification techniques by ruling out the existence of extra circuitry that can hide a hardware trojan with a test bypass. Thus, self-test techniques used in conjunction with IRIS can ensure the correct construction of security-critical hardware at all size scales.

## 1. Introduction

Cryptography tells us how to make a chain of trust rooted in special-purpose chips known as *secure elements*. But how do we come to trust our secure elements? Ideally, one can directly inspect the construction of a chip, but any viable inspection method must verify the construction of silicon chips *after* they have been integrated into finished products, *without* having to unmount or destroy the chips. The method should also ideally be *cheap and simple* enough for end users to access.

This paper introduces a technique we call "Infra-Red, *In-Situ*" (IRIS) inspection. It is founded on two insights: first, that silicon is transparent to infra-red light; second, that a digital camera can be modified to "see" in infra-red, thus effectively "seeing through" silicon chips. We can use these insights to inspect an increasingly popular family of chip packages known as Wafer Level Chip Scale Packages (WLCSPs) by shining infrared light through the back side of the package and detecting reflections from the lowest layers of metal using a digital camera. This technique works even after the chip has been assembled into a finished product. However, the resolution of the imaging method is limited to micron-scale features.

In this paper we will briefly review why silicon inspection is important as well as some current methods for inspecting silicon. Then, we will go into the IRIS inspection method, giving background on the theory of operation while disclosing methods and initial results. Finally, we'll contextualize the technique and discuss methods for closing the gap between micron-scale feature inspection and the nanometer-scale features found in today's chip fabrication technology.

### 1.1 Side Note on Trust Models

Many assume the point of trustable hardware is so that a third party can control what you do with your computer – like the secure enclave in an iPhone or a TPM in a PC. In this model, users delegate trust to vendors, and vendors do not trust users with key material: anti-tamper measures take priority over inspectability.

Readers who make this assumption would be confused by a trust method that involves open source and user inspections. To be clear, the threat model in this paper assumes no third parties can be trusted, especially not the vendors. The IRIS method is for users who want to be empowered to manage their own key material. The author acknowledges this is an increasingly minority position.

## 2. Why Inspect Chips?

The problem boils down to chips being literal black boxes with nothing but the label on the outside to identify them.

Figure 1 is an example of a survey done of the internal construction of microSD cards, from (Huang 2010). The survey was performed in an effort to trace down the root cause of a failed lot of products. Although every microSD card ostensibly advertised the same product and brand (Kingston 2GB), a decap study (where the exterior black epoxy is dissolved using a strong acid revealing the internal chips while destroying the card) revealed a great diversity in internal construction and suspected ghost runs. The take-away is that labels can't be trusted; if you have a high-trust situation, something more is needed to establish a device's internal construction than the exterior markings on a chip's package.

### 2.1 What Are The Existing Options for Inspecting Chips?

There are many options for inspecting the construction of chips; however, all of them suffer from a "Time Of Check versus Time Of Use" (TOCTOU) problem. In other words, none of these techniques are *in situ*. They must be performed either on samples of chips that are merely representative of the exact device in your possession, or they must be done at remote facilities such that the sample passes through many stranger's hands before returning to your possession.

Scanning Electron Microscopy (SEM), illustrated in Figure 2, is a popular method for inspecting chips. The technique can produce highly detailed images of even the latest nanometer-scale transistors. However, the technique is destructive: it can only probe the surface of a material. In order to image transistors one has to remove (through etching or polishing) the overlying layers of metal. Thus, the technique is not suitable for *in situ* inspection.

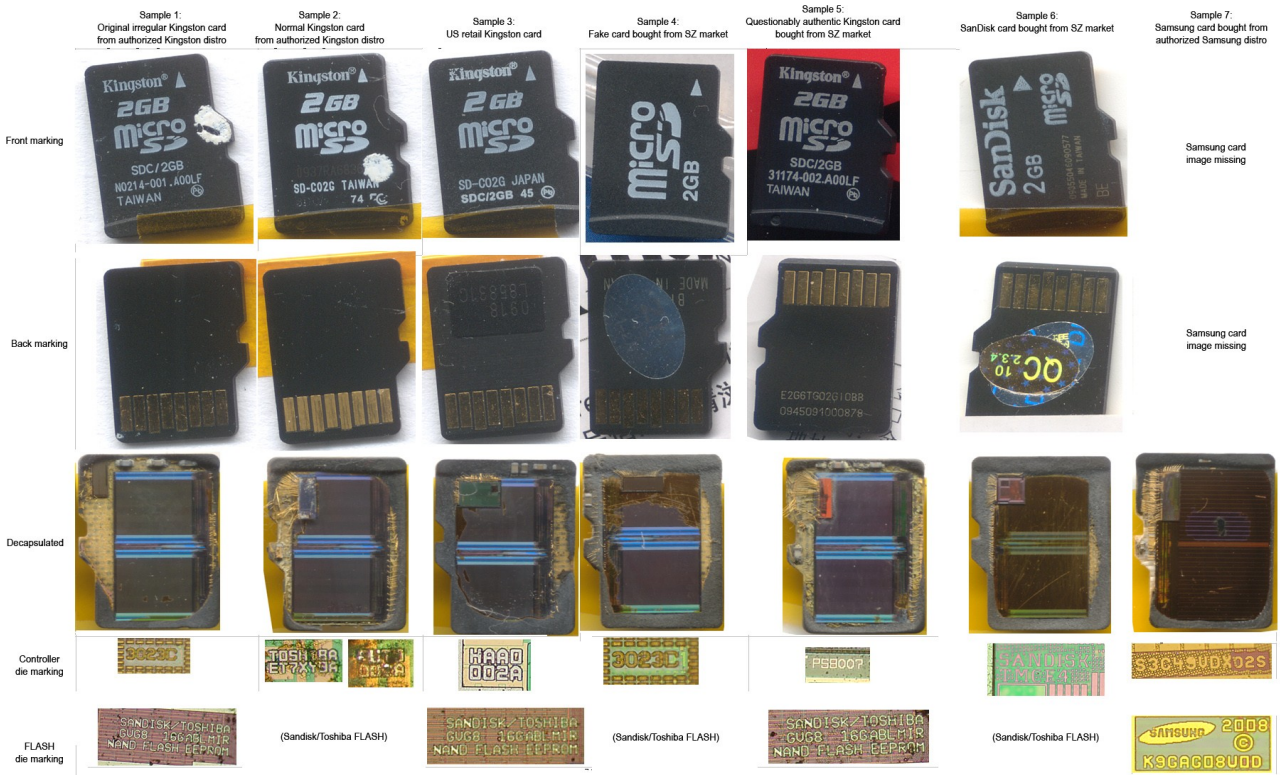

*Figure 1: A survey of microSD card internal construction, conducted in (Huang, 2017 pp. 164-165).*

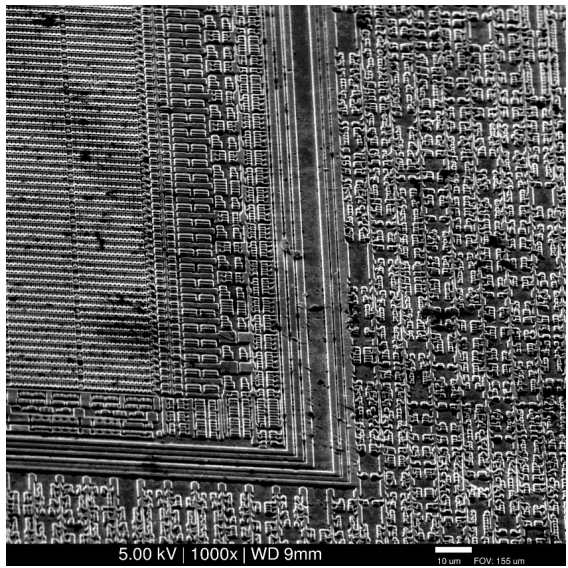

*Figure 2: Example SEM image of a silicon chip. Here individual transistors can be clearly seen. The upper metal layers have been removed prior to imaging by etching with a solution of hydrofluoric acid (Domke, 2023).*

X-rays are capable of non-destructive in-situ inspection (Figure 3); anyone who has traveled by air is familiar with the applicability of X-rays to detect foreign objects inside locked suitcases. However, silicon is nearly transparent to the types of X-rays used in security checkpoints, making it less suitable for establishing the contents of a chip package. It can identify the size of a die and the position of bond wires, but it can't establish much about the pattern of transistors on a die.

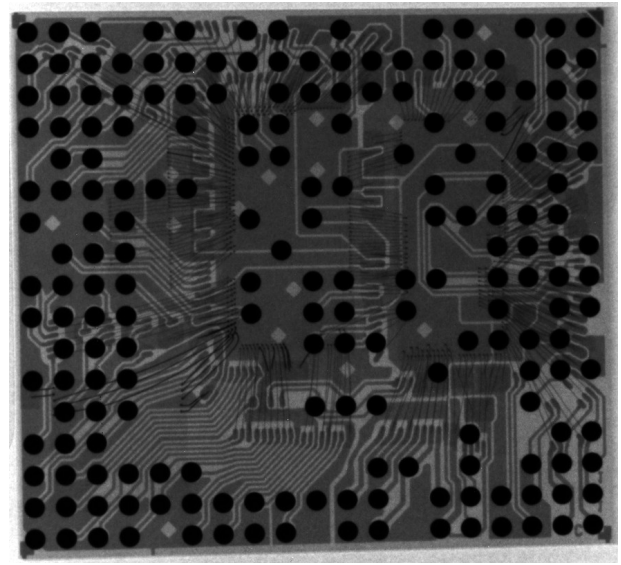

*Figure 3: An X-ray of a chip (Huang, 2017 p. 310)*

X-Ray Ptychography is a technique using high energy X-rays that can non-destructively establish the pattern of transistors on a chip (Figure 4). It is a very powerful technique, but unfortunately it requires a light source the size of a building, of which there are few in the world (Figure 5). While it is a powerful method, it is impractical for inspecting every end user device. It also suffers from the TOCTOU problem in that your sample has to be mailed to the Swiss Light Source (SLS) and then mailed back to you. So, unless you hand-carried the sample to and from the SLS, your device is now additionally subject to "evil courier" attacks.

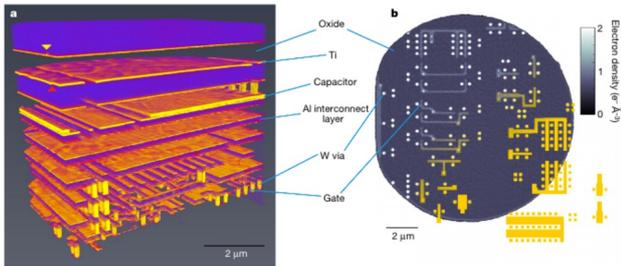

*Figure 4: Example Ptychographic X-ray Computed Tomography (PXCT). Adapted from (Holler, 2017).*

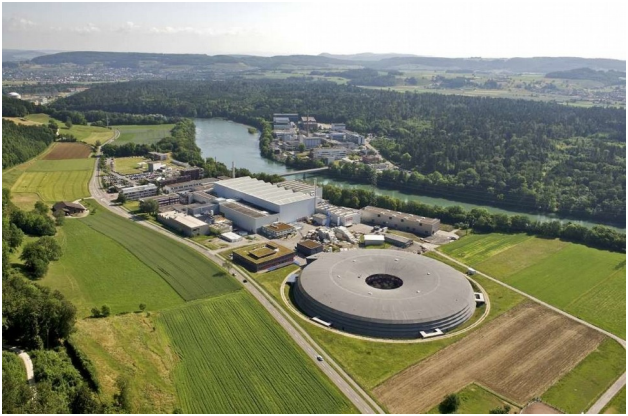

*Figure 5: Aerial view of the Swiss Light Source (SLS), used in (Holler, 2017). The donut-shaped building is the "lightbulb" needed to perform PXCT. (PSI, 2016).*

Optical microscopy – with a simple benchtop microscope, similar to those found in grade-school classrooms around the world – is also a noteworthy tool for inspecting chips that is easier to access than the SLS. Visible light can be a useful tool for checking the construction of a chip, if the chip itself has not been obscured with an opaque, over-molded plastic shell.

Fortunately, in the world of chip packaging, it has become increasingly popular to package chips with no overmolded plastic. The downside of exposing delicate silicon chips to possible mechanical abuse is offset by improved thermal performance, better electrical characteristics, smaller footprints, as well as typically lower costs when compared to overmolding. Because of its compelling advantages this style of packaging is ubiquitous in mobile devices. A common form of this package is known as the "Wafer Level Chip Scale Package" (WLCSP), and it can be optically inspected prior to assembly.

Figure 6 is an example of such a package viewed with an optical microscope, prior to attachment to a circuit board. In this image, the back side of the wafer is facing away from us, and the front side is dotted with 12 large silvery circles that are solder balls. The spacing of these solder balls is just 0.5mm – this chip would easily fit on your pinky nail.

The imaged chip is laying on its back, with the camera and light source reflecting light off of the top level routing features of the chip (Figure 7). Oftentimes these top level metal features take the form of a regular waffle-like grid. This grid of metal distributes power for the underlying logic, obscuring it from direct optical inspection.

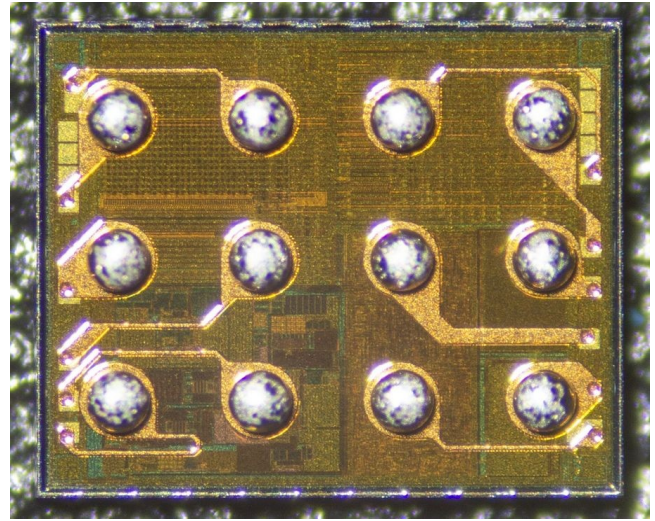

*Figure 6: A small WLCSP package, prior to assembly, imaged with visible light.*

Note that the terms "front" and "back" are taken from the perspective of the chip's designer; thus, once the solder balls are attached to the circuit board, the "front side" with all the circuitry is obscured, and the plain silvery or sometimes paint-coated "back side" is what's visible.

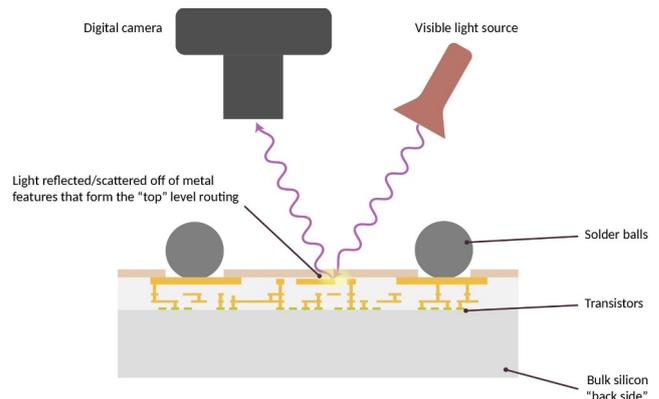

*Figure 7: Cross-section of the setup used to create the image in Figure 6.*

As a result, these chip packages look like opaque silvery squares, as demonstrated in Figure 8. Therefore front-side optical microscopy is not suitable for "in-situ" inspection, as the chip must be removed from the board in order to see the interesting bits on the front side of the chip.

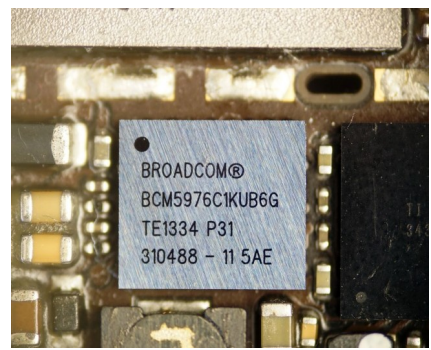

*Figure 8: A large WLCSP package, after attachment to a circuit board.*

## 2.2 The IRIS Inspection Method

The Infra-Red, In-Situ (IRIS) inspection method is capable of *seeing through* a chip already attached to a circuit board and non-destructively imaging the construction of a chip's logic.

## 2.3 Theory of Operation

Silicon goes from opaque to transparent in the range of 1000 nm to 1100 nm (shaded band in Figure 9). Above 1100 nm, it's as transparent as a pane of glass; below 1000 nm, it rapidly becomes more opaque than the darkest sunglasses.

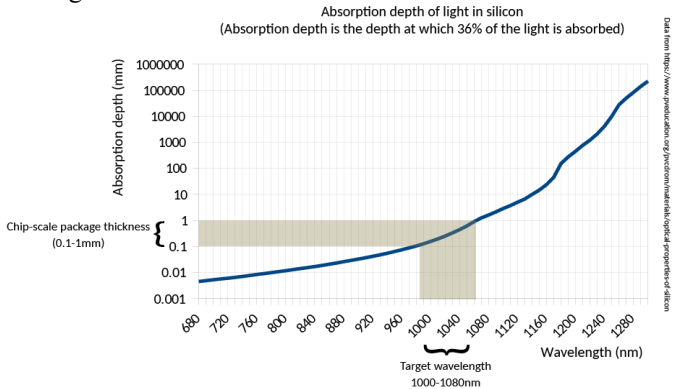

*Figure 9: Absorption depth of light versus wavelength. Adapted from (PVEducation, 2023).*

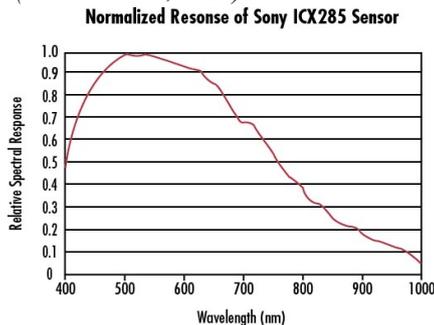

*Figure 10: A representative silicon sensor response curve. Adapted from (Edmunds 2023).*

Meanwhile, silicon-based image sensors retain some sensitivity in the near-to-short wave IR bands, as illustrated in Figure 10.

Between the curves in figures 9 and 10, there is a "sweet spot" where standard CMOS sensors retain some sensitivity to short-wave infrared, yet silicon is transparent enough that sufficient light passes through the layer of bulk silicon that forms the back side of a WLCSP package to do reflected-light imaging. More concretely, at 1000 nm a CMOS sensor might have 0.1x its peak sensitivity, and a 0.3 mm thick piece of silicon may pass about 10% of the incident light – so overall we are talking about a ~100x reduction in signal intensity compared to visible light operations. While this reduction is non-trivial, it is surmountable with a combination of a more intense light source and a longer exposure time (on the order of several seconds).

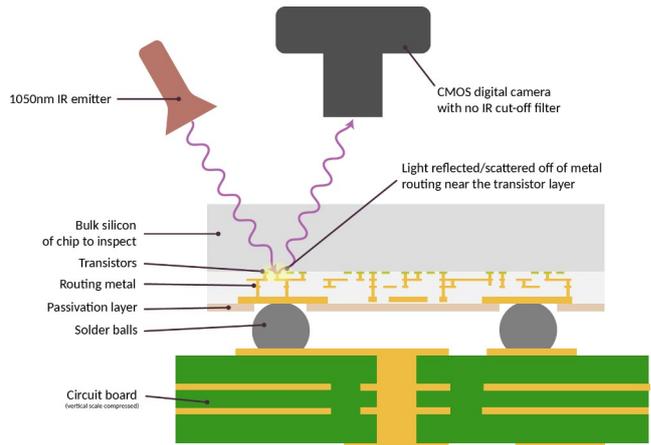

*Figure 11: Cross-section schematic of IRIS imaging.*

## 2.4 Implementation

Figure 11 is a cross-section schematic of the IRIS inspection setup. Here, the sample for inspection is already attached to a circuit board and we are shining light through the back side of the silicon chip. The light reflects off of the layers of metal closest to the transistors, and is imaged using a camera. Conceptually, it is fairly straightforward once aware of the "sweet spot" in infrared.

Two things need to be prepared for the IRIS imaging technique. First, the "IR cut-off filter" has to be removed from a digital camera. Normally, the additional infrared sensitivity of CMOS sensors is considered to be problematic, as it introduces color fidelity artifacts. Because of this excess sensitivity, all consumer digital cameras ship with a special filter installed that blocks any incoming IR light. Removing this filter can range from trivial to very complicated, depending on the make of the camera.

Second, we need a source of IR light. Incandescent bulbs and natural sunlight contain plenty of IR light, but our demonstration setup uses a pair of 1070 nm, 100 mA $I_F$ LED emitters from Martech, connected to a simple variable current power supply.

The inspiration for IRIS came from (Lohrke, 2018) and (Fritz, 2023). In (Lohrke 2018 this paper), a Phemos-1000 system by Hamamatsu (a roughly million dollar tool) uses a scanning laser to do optical backside imaging of an FPGA in a flip-chip package. More recently, (Frits 2023) demonstrates a similar technique, but using a much cheaper off-the-shelf Sony NEX-5T.

The images shown in this paper were taken using either a lens and camera assembly from (Hayear, 2023), where the IR cut-off filter was removed by the author, or a Sony A6000 camera modified by (Kolari, 2023) and mated with optics from (Omax, 2023). Instructions on how to remove the IR cut-off filter from (Hayear, 2023) can be found in (Huang, 2023).

## 2.5 Results (Sample Images)

Figures 12 & 13 shows an image of the larger WLCSP chip attached to a circuit shown in Figure 8, but taken in 1070 nm infrared light with the IRIS method.

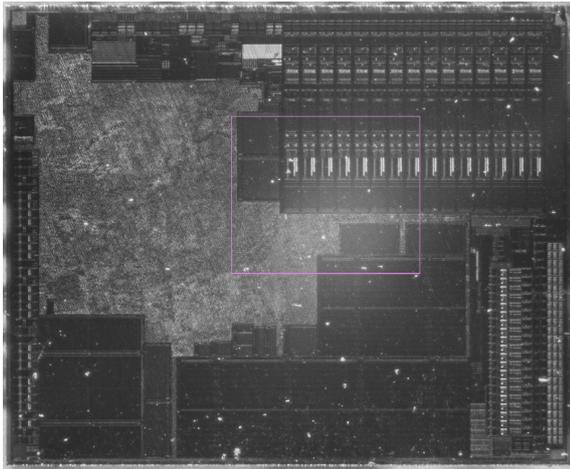

*Figure 12: The WLCSP chip in figure 8 imaged in infrared light.*

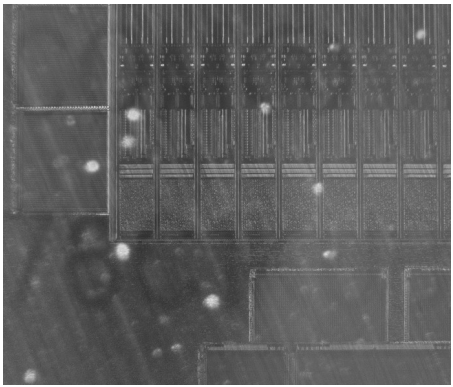

*Figure 13: the highlighted region in figure 9, at a higher magnification.*

The chip from Figures 8, 12 and 13 is the BCM5976, a capacitive touchscreen driver for older models of iPhones. Figure 12 clearly demonstrates the macro-scopic structure of the chip, with multiple channels of data converters on the top right and right edge, along with several arrays of non-volatile memory and RAM along the lower half. From the top left extending to the center is a sea of standard cell logic, which has a "texture" based on the routing density of the metal layers. Remember, we're looking through the backside of the chip, so the metal layer we're seeing is mostly M1 (the metal connecting directly to the transistors). The diagonal artifacts apparent through the standard cell region are due to a slight surface texture left over from wafer processing. Figure 13 demonstrates the imaging of meso-scopic structures, such as the row and structure column of memory macros and details of the data converters.

Figure 12 is 2330 pixels wide, while the chip is 3.9 mm wide: so each pixel corresponds to about 1.67 micron. To put that in perspective, if the chip were fabricated in 28 nm that would correspond to a "9-track" standard cell logic gate being 0.8 microns tall (based on data from (Wikichip, 2023)). Thus while this image cannot precisely resolve individual logic gates, the overall brightness of a region will bear a correlation to the type of logic gate used.

Figure 14 is an Armada610 chip, photographed in visible light. It is fabricated in a 55 nm process and packaged in a flip-chip BGA (FCBGA). FCBGA is a popular package type, but more importantly for IRIS the silicon is pre-thinned and mirror-polished. This is done to improve thermal performance, but also makes for very clean backside images.

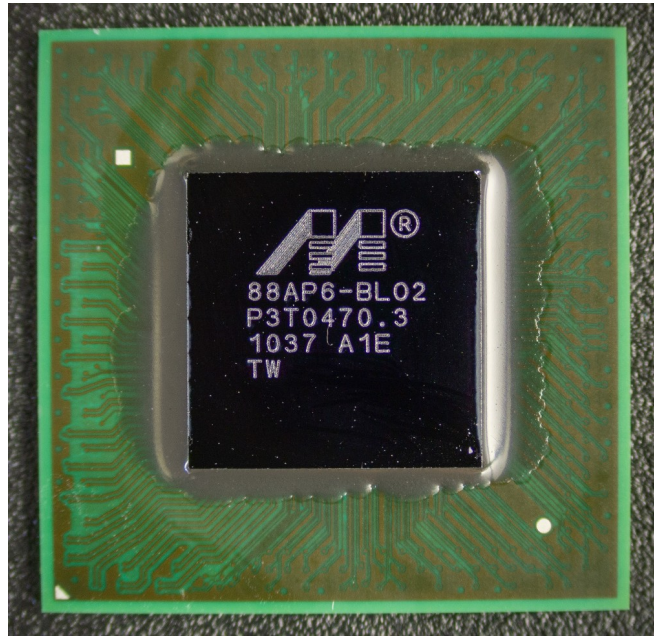

*Figure 14: Sample chip in visible light.*

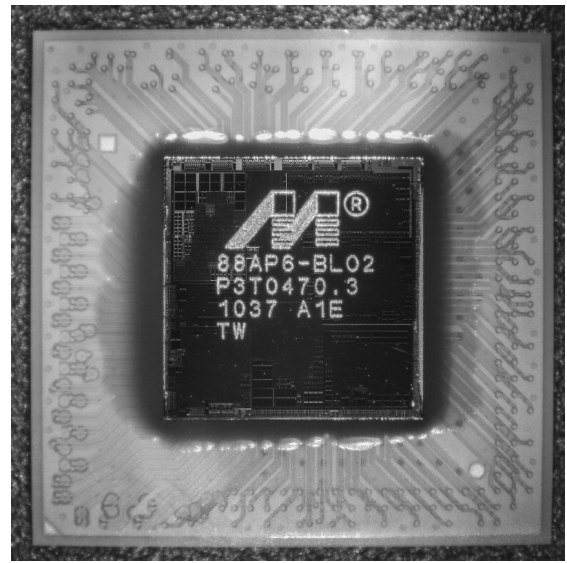

*Figure 15: The chip in figure 14, but imaged in 1070 nm IR. The light source is shining from the top right. Some chip details are already visible. The die is 8mm wide.*

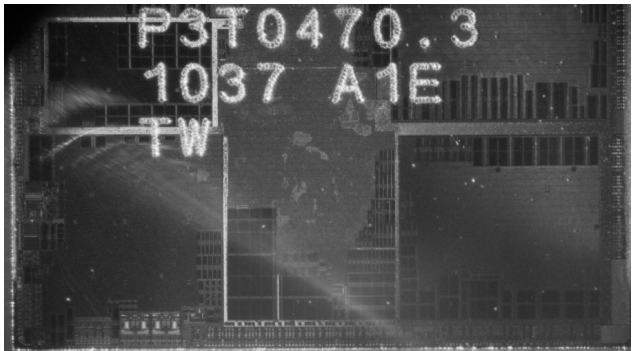

*Figure 16: Lower portion of figure 15.*

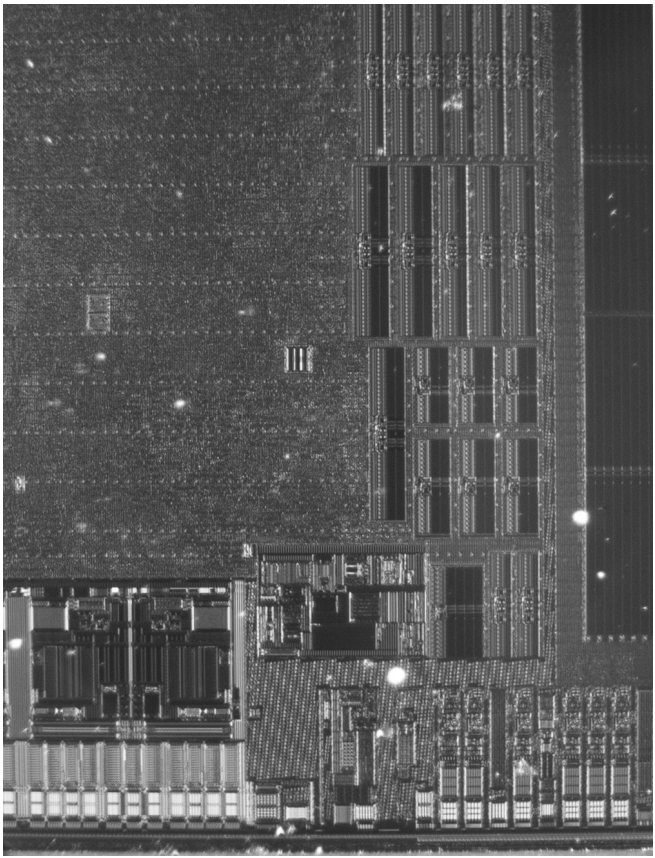

*Figure 17: magnified region of figure 16, from just next to the lower left hand corner.*

Figure 16 is the lower part of the chip in figure 14, but imaged in 1070 nm IR. Here we can start to clearly make out the shapes of memory macros, I/O drivers, and regions of differing routing density in the standard cell logic. The die is about 4290 pixels across in the full-resolution image – about 1.86 microns per pixel.

Figure 17 is higher magnification of a region towards the lower left edge of Figure 16. Here we can make out the individual transistors used in I/O pads, sense amps on the RAM macros, and the texture of the standard cell logic. The resolution of this photo is roughly 1.13 microns per pixel – around the limit of what could be resolved with the 1070 nm light source – and a hypothetical "9-track" standard cell logic gate would be a little over a pixel tall by a couple pixels wide.

## 3. Discussion

IRIS inspection reveals the internal structure of a silicon chip. IRIS can do this "in situ" (after the chip has been assembled into a product), and in a non-destructive manner. However, the technique can only inspect chips that have been packaged with the back side of the silicon exposed. Fortunately, a fairly broad and popular range of packages such as WLCSP and FCBGA already expose the back side of chips.

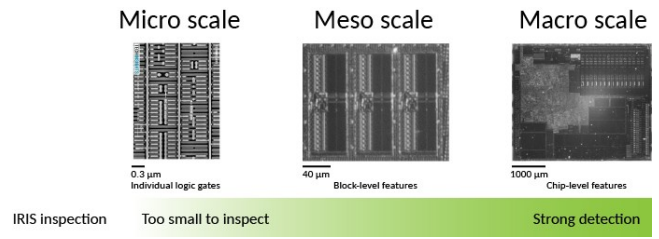

*Figure 18: Various size scales found on a chip, in relationship to IRIS capabilities.*

IRIS cannot inspect the smallest features of a chip. Figure 18 illustrates the various size scales found on a chip and relates it to the capabilities of IRIS. The three general feature ranges are prefixed with micro-, meso-, and macro-. On the left hand side, "micro-scale" features such as individual logic gates will be smaller than a micron tall. These are not resolvable with infra-red wavelengths and as such not directly inspectable via IRIS, so the representative image was created using SEM. The imaged region contains about 8 individual logic gates.

In the middle, we can see that "meso-scale" features can be constrained in size and identity. The representative image, taken with IRIS, shows three RAM "hard macros" in a 55 nm process. Individual row sense amplifiers are resolvable in this image. Even in a more modern sub-10 nm process, we can constrain a RAM's size to plus/minus a few rows or columns.

On the right, "macro-scale" features are clearly enumerable. The number and count of major functional blocks such as I/O pads, data converters, oscillators, RAM, FLASH, and ROM blocks are readily identified.

IRIS is a major improvement over simply reading the numbers printed on the outside of a chip's package and taking them at face value. It's comparable to being able to X-ray every suitcase for dangerous objects, versus accepting suitcases based solely on their exterior size and shape.

Even with this improvement, malicious changes to chips – referred to as "hardware trojans" – can in theory remain devilishly difficult to detect, as demonstrated in (Becker, 2013). This paper proposes hardware trojans that only modulate the doping of transistors. Doping modifications would be invisible to most forms of inspection, including SEM, X-Ray ptychography, and IRIS.

The good news is that the attacks discussed in (Becker, 2013) are against targets that are entirely unhardened against hardware trojans. With a reasonable amount of design-level hardening, we may be able to up the logic footprint for a hardware trojan into something large enough to be detected with IRIS. Fortunately, there is an existing body of research on hardening chips against trojans, using a variety of techniques including logic locking, built in self test (BIST) scans, path delay fingerprinting, and self-authentication methods (Tehranipoor, 2014).

IRIS is a necessary complement to logic-level hardening methods, because logic-only methods are vulnerable to bypasses and emulation. In this scenario, a hardware trojan includes extra circuitry to evade detection by spoofing self-tests with correct answers, like a wolf carrying around a sheep's costume that it dons when a shepherd is nearby. Since IRIS can constrain meso-scale to macro-scale structure, we can rule out medium-to-large scale circuit modifications, giving us more confidence in the results of the micro-scale verification as reported by logic-level hardening methods.

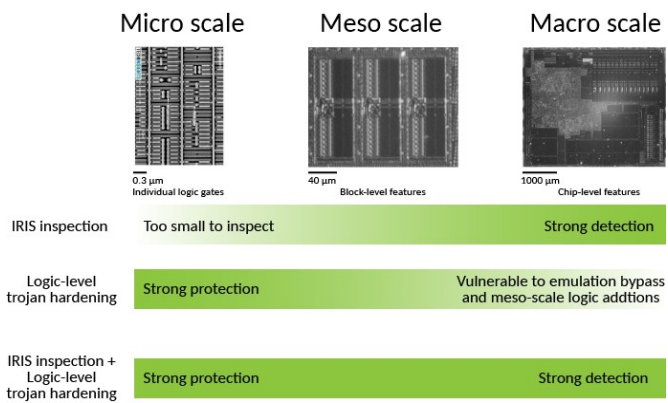

*Figure 19: Comparison of the detection-vs-protection trade offs of logic level hardening and IRIS inspection.*

Thus, IRIS can be used in conjunction with logic-level trojan hardening to provide an overall high-confidence solution in a chip's construction using non-destructive and in-situ techniques, as illustrated in Figure 19.

The primary requirement of the logic-level hardening method is that it must not be bypassable with a trivial amount of logic. For example, simple "logic locking" (a method of obfuscating logic which in its most basic form inserts X(N)ORs in logic paths, requiring a correct "key" to be applied to one input of the X(N)ORs to unlock proper operation) could be bypassed with just a few gates once the key is known, so this alone is not sufficient. However, a self-test mechanism that blends state from "normal runtime" mode and "self test" mode into a checksum of some sort could present a sufficiently high bar. In such a stateful verification mechanism, the amount of additional logic required to spoof a correct answer is proportional to the amount of state accumulated in the test. Thus, one can "scale up" the coverage of a logic-level test by including more state, until the point where any reliable bypass would be large enough to be detected by IRIS. The precise amount of state would depend on the process geometry: smaller process geometries would need more state.

Under the assumption that each extra bit would imply an additional flip flop plus a handful of gates, a back-of-the-envelope calculation indicates a 28 nm process would require just a few bits of state in the checksum. In this scenario, the additional trojan logic would modify several square microns of chip area, and materially change the scattering pattern of infra-red light off of the chip in the region of the modification. Additional techniques such as path delay fingerprinting may be necessary to force the trojan logic to be spatially clustered, so that the modification is confined to a single region, instead of diffused throughout the standard cell logic array.

## 4. Summary and Future Direction

IRIS is a promising technique for improving trust in hardware. With a bit of foresight and planning, designers can use IRIS in conjunction with logic hardening to gain comprehensive trust in a chip's integrity from micro- to macro-scale. While the technique may not be suitable for every chip in a system, it fits comfortably within the parameters of chips requiring high assurance such as trust roots and secure enclaves.

Of course, IRIS is most effective when combined with open source chip design. In closed source chips, we don't know what we're looking at, or what we're looking for; but with open source chips we can use the design source to augment the capabilities of IRIS to pinpoint features of interest.

We hope that IR-capable microscopes become a staple on hardware hacker's workbenches, so we can start to assemble databases of what chips *should* look like – be they open or closed source. Such a database can also find utility in everyday supply chain operations, helping to detect fake chips or silent die revisions prior to device assembly.

Over the coming year, we look to improve the core IRIS technique. In addition to upgrading optics and adding image stitching to our toolbox, digitally controlling the angle and azimuth of incident light should play a significant role in enhancing the utility of IRIS. The sub-wavelength features on a chip interact with incident light like a hologram. By modifying the angle of lighting, we can likely glean even more information about the structure of the underlying circuitry, even if they are smaller than the diffraction limit of the system.

A bit further down the road, we can see IRIS being combined with active laser probing techniques, where IRIS is used to precisely locate a spot that is then illuminated by an intense laser beam. While this has obvious applications in fault induction, it can also have applications in verification and chip readout. For example, the localized thermal stimulation of a laser can induce the Seeback effect, creating a data-dependent change in power consumption detectable with sensitive current monitors. We note here that if physical tamper-resistance is necessary, post-verification a chip can be sealed in opaque epoxy with bits of glitter sprinkled on top to shield it from direct optical manipulation attacks and evil-maid attacks. However, this is only necessary if these attacks are actually part of the threat model. Supply chain attacks happen, by definition, upstream of the end user's location.

The other half of optical chip verification is an image processing problem. It's one thing to have reference images of the chip, and it's another thing to be able to take the image of a chip and compare it to the reference image and generate a confidence score in the construction of the chip. A turnkey feature extraction and comparison tool would go a long way toward making IRIS a practically useful tool.

Ultimately, the hope is to create a verification solution that grows in parallel with the open source chip design ecosystem, so that one day we can have chips we can trust. Not only will we know what chips are intended to do, we can rest assured knowing they were built as intended, too.

## 5. Acknowledgements

This research is partially funded by a NGI Zero "Privacy & Trust Enhancing Technologies" (PET) grant from NLnet and the European Commission, as well as by the donations of Github Sponsors.

The author also thanks Dmitry Nedospadov, Felix Domke and Jannis Harder for their helpful conversations and direction.

## 6. Copyright

## 7. Bibliographical References